\documentclass[crystals,review,submit,moreauthors,dvipdfmx,12pt,a4paper]{mdpi}

\usepackage{amssymb,amsmath}
\usepackage{graphicx}

\def\nn{\nonumber}

\newcommand{\bsigma}{\mbox{\boldmath $\sigma$}}

\Title{The origin of Raman $D$ Band: Bonding and Antibonding Orbitals in Graphene}

\Author{Ken-ichi Sasaki~$^{1,\star}$, Yasuhiro Tokura~$^{1,2}$, and Tetsuomi Sogawa~$^{1}$}

\address{%
$^{1}$ NTT Basic Research Laboratories, Nippon Telegraph and Telephone
Corporation, 3-1 Morinosato Wakamiya, Atsugi, Kanagawa 243-0198,
Japan \\
$^{2}$ Graduate School of Pure and Applied Sciences,
University of Tsukuba, Tsukuba, Ibaraki 305-8571, Japan
}

\corres{sasaki.kenichi@lab.ntt.co.jp}

\abstract{
\nolinenumbers
In Raman spectroscopy of graphite and graphene, 
the $D$ band at $\sim 1355$cm$^{-1}$ 
is used as the indication of the dirtiness of a sample.
However, our analysis suggests that the physics behind 
the $D$ band is closely related to a very clear idea 
for describing a molecule, namely bonding and antibonding
orbitals in graphene.
In this paper, we review our recent work on 
the mechanism for activating the $D$ band at a graphene edge.
}


\begin{document}
%%%%%%%%%%%%%%%%%%%%%%%%%%%%%%%%%%%%%%%%%%%%%%%%%%%%%%%%%%%%
\nolinenumbers
\section{Introduction}

Bonding and antibonding orbitals are basic ideas 
for describing molecules.
Bonding orbitals contribute to the formation of a molecule, 
whereas antibonding orbitals weaken the bonding and 
destabilize a molecule.
Normally, bonding orbitals are more stable than 
antibonding orbitals in terms of energy and thus 
a molecule is stable unless 
sufficient electrons occupy the antibonding orbitals.

Graphene~\cite{novoselov05,zhang05} is unique 
with respect to its molecular orbitals.
The bonding and antibonding orbitals in graphene are degenerate, 
and various types of linear combination of these orbitals 
form the Fermi surface expressed by the isoenergy sections of Dirac
cones.
This degeneracy plays an essential role in various phenomena.
For example, graphene is stable
with respect to a large shift of the Fermi energy position~\cite{chen11}.
Another notable example is that graphene exhibits high mobility.
Elastic backward scattering between 
the bonding and antibonding orbitals
induced by long-range impurity
potential is suppressed because they are orthogonal~\cite{ando98}.
In this paper, we show that the bonding and antibonding orbitals in
graphene 
are key factors in the activation mechanism of the $D$ band observed at
a graphene edge.

Since the discovery of the $D$ band,
much interest has focused on its origin.
Tuinstra and Koenig attributed
the $D$ band to an $A_{1g}$ zone-boundary mode 
at the edge of a sample (see Fig.~\ref{fig:unit}) 
on the grounds that 
the Raman intensity is proportional to the edge percentage and 
that the edge causes a relaxation of the momentum conservation
needed for activating a zone-boundary phonon~\cite{tuinstra70}.
Katagiri et al. confirmed that the $D$ band originates from an edge 
(or discontinuity in the carbon network) by observing 
the light polarization dependence of the $D$ band intensity 
at graphite edge planes~\cite{katagiri88}.
The atomic arrangement of an edge has two principal axes;
armchair and zigzag edges.
Can\ifmmode \mbox{\c{c}}\else \c{c}\fi{}ado et al.
showed that the armchair (zigzag) edge is relevant (irrelevant) to 
the relaxation of the momentum conservation 
for a zone-boundary phonon~\cite{canifmmode04sec}.
In addition, they found that 
the $D$ band Raman intensity depends on the polarization of laser light,
that is, the intensity is maximum (minimum) 
when the polarization is parallel (perpendicular) to the armchair
edge.
The light polarization dependence of the $D$ band is also observed ubiquitously
at the armchair edges of a single layer of graphene, which suggests that 
out-of-plane coupling in graphite is not essential to the origin of
the $D$ band~\cite{you08,gupta09,cong10}.

A model of the $D$ band must at least explain the observed properties:
the $D$ band intensity increases only at an armchair edge and 
is dependent on the laser light polarization.
%In addition to these requirements, a good model should give new prediction.

The current $D$ band model is a double resonance model~\cite{thomsen00}.
In this model, a photo-excited electron passes through two resonance states,
which enhances the Raman intensity of a phonon with nonzero wave vector ${\bf
q}\ne 0$.
This model is not concerned with the details of
electron-phonon and electron-light matrix elements, 
and it does not provide clear explanations of the properties
of the $D$ band.
Also, the intensity calculated with this model
is dependent on the lifetime of the resonance states.
Usually, the lifetime is determined in such a manner that 
a calculated result reproduces experimental data. 
In this sense, the double resonance model is phenomenological.
Because the lifetime is shorter in a defective graphene sample, 
double resonance does not necessarily mean an enhancement of the $D$ band Raman intensity.
On the other hand, the model can account for the so-called dispersive
behavior of the $D$ band~\cite{thomsen00}.
However, as we will show in this paper, 
dispersive behavior is characteristic of $A_{1g}$ modes, 
rather than an inherent property of the $D$ band.
In fact, dispersive behavior is observed also for the $2D$
band~\cite{vidano81}, and the excitation does not need an edge, 
which is in contrast to the $D$ band.

In this paper we show that the observed properties of the $D$ band 
are naturally explained 
in terms of simple ideas based on molecular orbitals and momentum conservation.
In our formulation,
the $D$ band is excited from a photo-excited electron through a single
resonance process in the same way as the $G$ band.\footnote{
Negri et al. took the same approach to the resonance Raman process of
the $D$ band of polycyclic aromatic hydrocarbons~\cite{negri04}.}
It is concluded that, without invoking an artificial assumption, 
the $D$ band is closely related to 
(1) the orbital dependence of the electron-phonon matrix element,
(2) the special nature of the armchair edge, 
and (3) optical anisotropy~\cite{sasaki11_Dband}.

This paper is organized as follows.
In Sec.~\ref{sec:bab} we observe that 
a graphene molecular orbital and wave vector are closely correlated.
This correlation is both an important factor in terms of understanding
the $D$ band and an essential feature of graphene.
In Sec.~\ref{sec:main} 
the properties of the $D$ band are deduced from three factors (1,2,3).
The importance of the electron-phonon matrix element and the role of the
armchair edge in the excitation mechanism of the $D$ band  
are explained in detail.
In Sec.~\ref{sec:pre}
we show some predictions obtained with our model.
Future prospects and our conclusion are given in Sec.~\ref{sec:out}
and Sec.~\ref{sec:con}, respectively.
We give some notes on resonant condition in Appendix~\ref{app:rc}.

%%%%%%%%%%%%%%%%%%%%%%%%%%%%%%%%%%%%%%%%%%%%%%%%%%%%%%%%%%%%

\section{Bonding and antibonding orbitals}\label{sec:bab}

\begin{figure}[htbp]
 \begin{center}
  \includegraphics[scale=0.9]{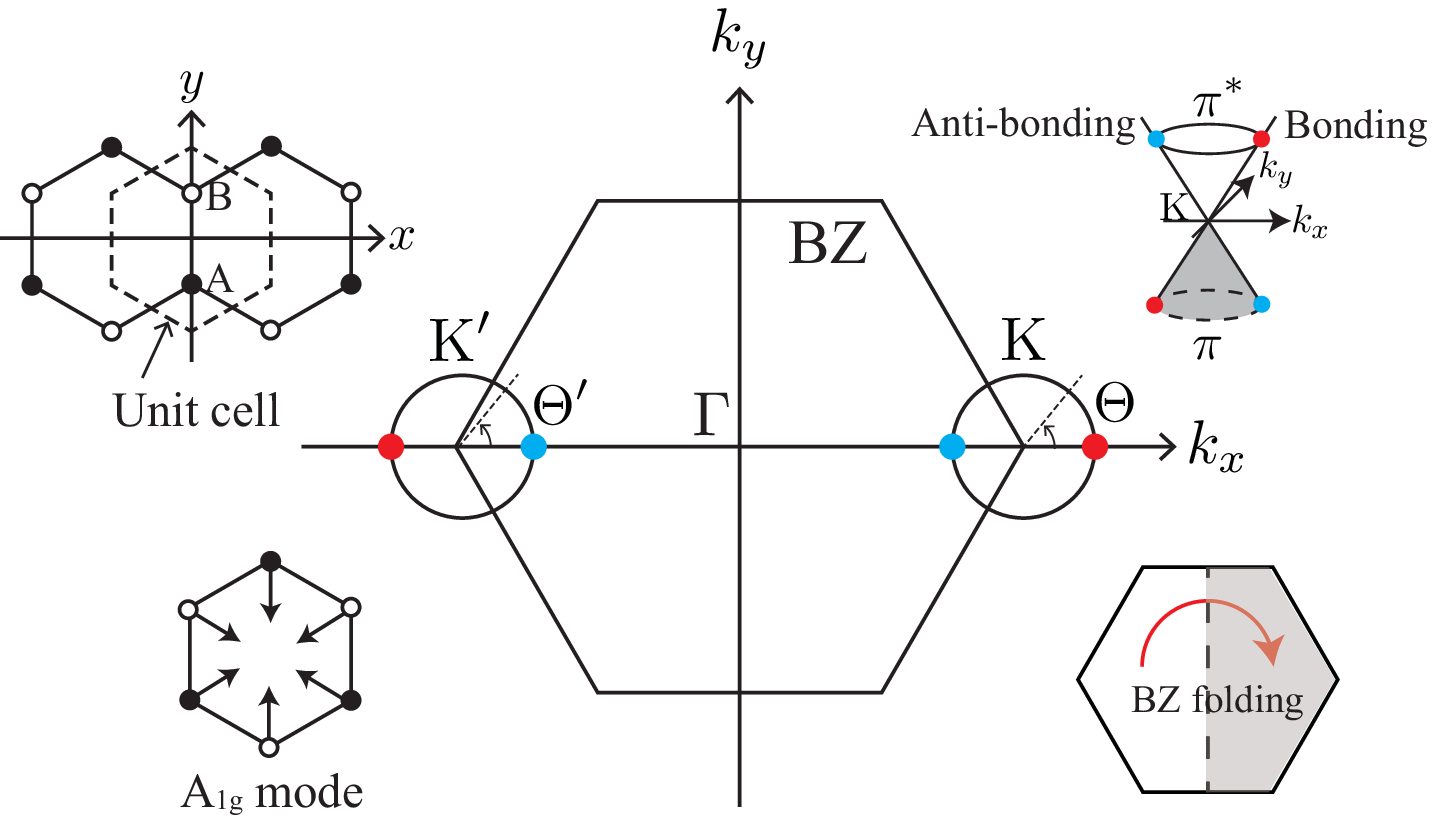}
 \end{center}
 \caption{Graphene unit cell and BZ.
 The K (K$'$) point is located at ${\bf k}_{\rm F}=(4\pi/3a,0)$ 
 ($-{\bf k}_{\rm F}$), where $a$ is the lattice constant.
 The positions of the bonding and antibonding orbitals are marked by red and
 blue points on the iso-energy sections of the Dirac cones (circles).
 The Raman $D$ band is composed of zone-boundary $A_{1g}$ modes
 that consist only of C-C bond stretching motions.
 The armchair edge identifies a state at $(k_x,k_y)$ with a state at
 $(-k_x,k_y)$, and causes the BZ folding.
 }
 \label{fig:unit}
\end{figure}

Graphene's hexagonal unit cell 
has two carbon atoms, denoted by A and B in Fig.~\ref{fig:unit}, 
and the electron's wave function $\psi$ is written as
a linear combination of $2p_z$ atomic orbitals 
of the A and B atoms, $\chi_{\rm A}$ and $\chi_{\rm B}$.
When we apply Bloch's theorem to graphene, 
we obtain $\psi$ and the band structure 
as a function of the wave vector ${\bf k}$~\cite{wallace47,slonczewski58}.
In the Brillouin zone (BZ) of graphene, 
there are two points, namely the K and K$'$ points, where
the conduction and valence bands touch each other. 
The orbitals of the states near the K point take the form of
\begin{align}
 \psi_{\rm K}^s({\bf k}) = 
 \frac{1}{\sqrt{2}} \left(
 e^{-i\Theta({\bf k})}\chi_{\rm A} + s \chi_{\rm B} \right),
 \label{eq:wf1}
\end{align}
where the phase $\Theta({\bf k}) \in [0,2\pi]$
is the polar angle between vector ${\bf k}$ measured
from the K point and the $k_x$-axis (see Fig.~\ref{fig:unit}),
and $s=\pm1$ is the band index ($s=+1$ is the $\pi^*$-band and $s=-1$
$\pi$-band).
The orbitals of the states near the K$'$ point are written as 
\begin{align}
 \psi_{\rm K'}^s({\bf k}) = \frac{1}{\sqrt{2}} \left(
  -e^{i\Theta'({\bf k})}\chi_{\rm A} + s \chi_{\rm B} \right),
 \label{eq:wf2}
\end{align}
where $\Theta'({\bf k})$ is the polar angle defined with respect to the K$'$ point 
as shown in Fig.~\ref{fig:unit}.

%%%%%%%%%%%%%%%%%%%%%%%%%%%%%%%%%%%%%%%%%%%%%%%%%%%%%%%%%%%%
%%%%%%%%%%%%%%%%%%%%%%%%%%%%%%%%%%%%%%%%%%%%%%%%%%%%%%%%%%%%

In Eq.~(\ref{eq:wf1}),
the bonding and antibonding orbitals 
($\chi_{\rm A}+\chi_{\rm B}$ and $-\chi_{\rm A}+\chi_{\rm B}$)
are located at $\Theta=0$ and $\pi$, respectively, 
on the iso-energy section of the $\pi^*$-band ($s=+1$).
The orbital with a general $\Theta$ 
is a linear combination of the bonding and antibonding orbitals.
In Eq.~(\ref{eq:wf2}), 
the bonding and antibonding orbitals 
are located at $\Theta'=\pi$ and $0$, respectively,
on the iso-energy section of the $\pi^*$-band.
As we can see in Fig.~\ref{fig:unit}, 
the bonding (antibonding) orbitals are located symmetrically on the
$k_x$-axis with respect to the $\Gamma$ point,
which is one of the most important characteristics of the BZ of
graphene [i.e., mirror symmetry with respect to the replacement, 
$x \leftrightarrow -x$].

The orbitals Eqs.~(\ref{eq:wf1}) and (\ref{eq:wf2}) 
can be derived from an effective model for graphene, 
that is, a massless Dirac equation, in the following manner.
The energy eigenequation is written as
\begin{align}
 E \psi = 
 \hat{H} \psi=
 v_{\rm F} 
 \begin{pmatrix}
  \bsigma \cdot \hat{\bf p} & 0 \cr
  0 & \bsigma' \cdot \hat{\bf p}
 \end{pmatrix}
 \begin{pmatrix}
  \psi_{\rm K}({\bf r}) \cr 
  \psi_{\rm K'}({\bf r}) 
 \end{pmatrix},
 \label{eq:H0}
\end{align}
where 
$v_{\rm F}$ is the Fermi velocity,
$\hat{\bf p}=(\hat{p}_x,\hat{p}_y)$ is a momentum operator,
$0$ in the off-diagonal terms represents a $2\times 2$ null matrix,
and $\bsigma=(\sigma_x,\sigma_y)$ and $\bsigma'=(-\sigma_x,\sigma_y)$
are $2\times 2$ Pauli spin matrices:
\begin{align}
 \sigma_x = 
 \begin{pmatrix}
  0 & 1 \cr 1 & 0
 \end{pmatrix}, \ \ \ 
 \sigma_y = 
 \begin{pmatrix}
  0 & -i \cr i & 0
 \end{pmatrix}.
\end{align}
The massless Dirac equation in Eq.~(\ref{eq:H0}) is decomposed into two
Weyl's equations for the K and K$'$ valleys: $E\psi_{\rm K}({\bf r}) = v_{\rm F}\bsigma \cdot \hat{\bf p}\psi_{\rm K}({\bf r})$
and $E\psi_{\rm K'}({\bf r}) =v_{\rm F}\bsigma' \cdot \hat{\bf
p}\psi_{\rm K'}({\bf r})$, respectively.
To reproduce Eq.~(\ref{eq:wf1}), we assume the plane wave solution 
$\psi_{\rm K}({\bf r})= e^{i{\bf k}\cdot {\bf r}} \psi_{\rm K}({\bf k})/\sqrt{V}$,
and obtain the energy eigenequation $\hbar v_{\rm F} \bsigma\cdot {\bf k}\psi_{\rm
K}({\bf k})=E \psi_{\rm K}({\bf k})$.
Then, the corresponding energy eigenvalue and eigenstate are easy to find 
using $k_x \mp ik_y=|{\bf k}|e^{\mp i\Theta({\bf k})}$ as $E=s \hbar
v_{\rm F}|{\bf k}|$ and 
\begin{align}
 \psi_{\rm K}^s({\bf k}) = \frac{1}{\sqrt{2}}
 \begin{pmatrix}
  e^{-i\Theta({\bf k})} \cr s
 \end{pmatrix}.
 \label{eq:wfK}
\end{align}
This is identical to Eq.~(\ref{eq:wf1}) by setting 
\begin{align}
 \chi_{\rm A} =
 \begin{pmatrix}
 1 \cr 0
 \end{pmatrix}, \ \ \  
 \chi_{\rm B} =
 \begin{pmatrix}
 0 \cr 1
 \end{pmatrix}.
\end{align}
Similarly, we can check that Eq.~(\ref{eq:wf2}) is the solution of 
$\hbar v_{\rm F} \bsigma'\cdot {\bf k}\psi_{\rm K'}({\bf k})=E \psi_{\rm
K'}({\bf k})$, which is given by
\begin{align}
 \psi_{\rm K'}^s({\bf k}) = \frac{1}{\sqrt{2}}
 \begin{pmatrix}
  -e^{i\Theta'({\bf k})} \cr s
 \end{pmatrix}.
\end{align}

The Dirac equation is very helpful as regards understanding the mechanism of a
result in terms of symmetry and momentum conservation.
In particular, the fact that the Dirac equation is composed of 
a multiplication of the Pauli matrices (for the A and B atoms) and 
momentum operators makes it easy to recognize that
the orbital (the pattern of the linear combination of $\chi_{\rm A}$ and
$\chi_{\rm B}$) is dependent on the wave vector of the particle.
In the following, we refer to the Dirac equation for graphene
in order to capture the essential features of a result.
The graphene Dirac equation differs from the original Dirac equation
in the following respect:
the wave function $\psi$ in the graphene Dirac equation
has 4 components consisting of two orbitals ($\chi_{\rm A}$ and $\chi_{\rm B}$)
and two valleys (K and K$'$), while those in the original Dirac equation
are two spin states (up and down spins) and two chiralities
(left- and right-handed)~\cite{sakurai67}.
Thus, the orbital degree of graphene is commonly referred to as pseudo-spin.\footnote{
We define the bonding and antibonding orbitals by the molecular orbitals
of nearest neighbor atoms having the same position in the $x$-axis (as
shown in Fig.~\ref{fig:unit}). Although our definition is not appropriate for the
bonds between nearest neighbor atoms having different positions in the
$x$-axis, those are not so important in discussing the $D$ band at armchair edge~\cite{sasaki11_Dband}.
Note also that our definition connects smoothly to the definition of
$\pi$ and $\pi^*$ bands at the $\Gamma$ point, where any nearest
neighbor carbon atoms form the bonding (antibonding) orbital in the $\pi$ ($\pi^*$) band.
}

\section{Light polarization dependence of $D$ band intensity}\label{sec:main}

In this section we show that the light polarization dependence of the
$D$ band originates from three factors. 
The first factor concerns the nature of the electron-phonon interaction
for the $A_{1g}$ mode, which will be explained in Sec.~\ref{ssec:back}.
The second factor is a modification of the BZ by the armchair edge, 
which will be explored in Sec.~\ref{ssec:bzh}.
In Sec.~\ref{ssec:opan}, we describe the third factor, which concerns
the interaction between electrons and a polarized laser light.
In Sec.~\ref{ssec:fin}, we construct the $D$ band polarization formula,
by combining the three factors.

\subsection{Dominance of intervalley backward scattering}\label{ssec:back}

%%%%%%%%%%%%%%%%%%%%%%%%%%%%%
\begin{figure}[htbp]
 \begin{center}
  \includegraphics[scale=0.9]{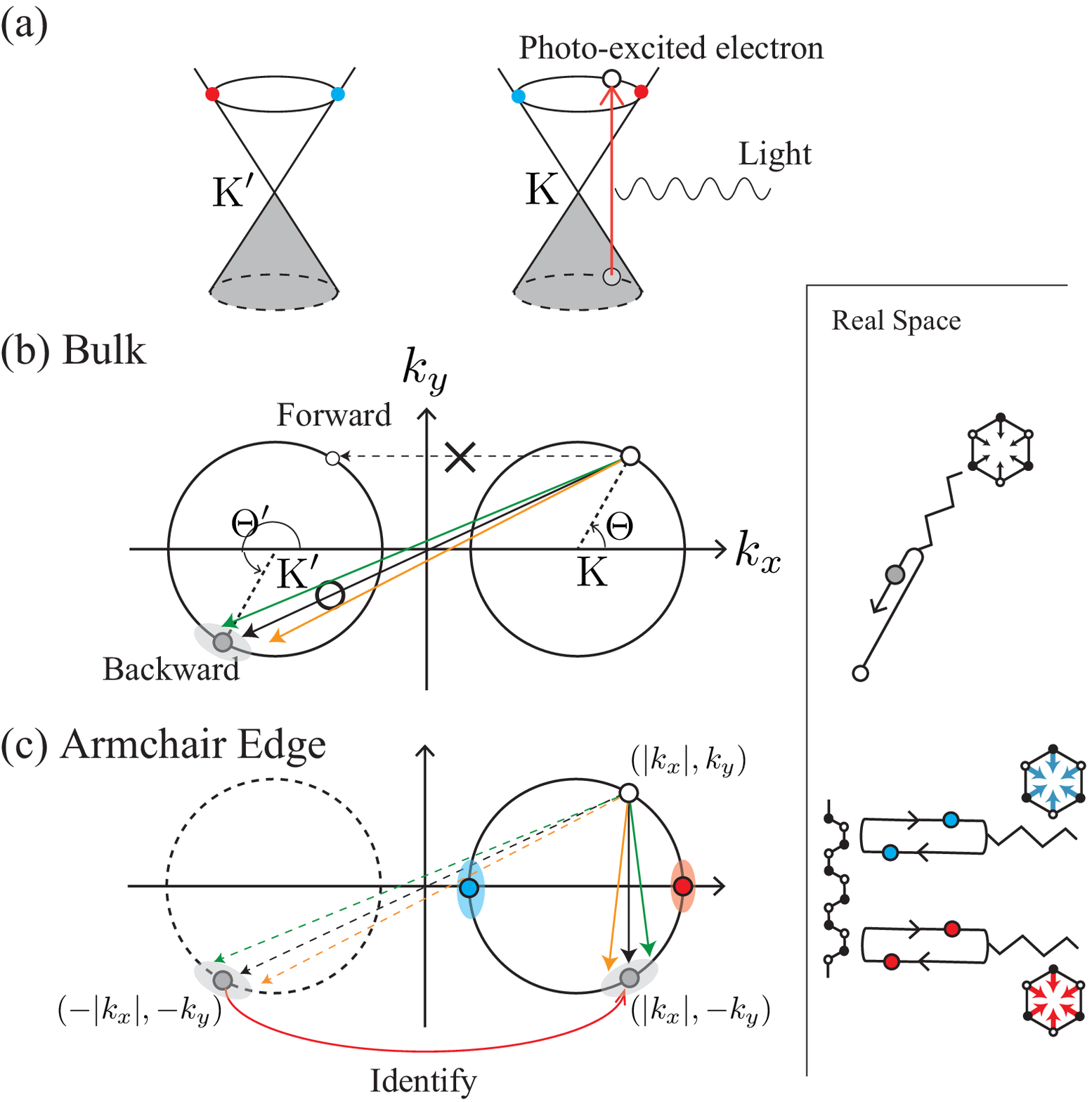}
 \end{center}
 \caption{Mechanism of $D$ band intensity enhancement at armchair edge.
 (a) A photo-excited electron near the K point.
 (b) The electron-phonon interaction of an zone-boundary $A_{1g}$ mode 
 results in the dominance of intervalley backward scattering.
 (c) 
 The BZ folding leads to the appearance of two special electronic states
 (with the bonding [red] and antibonding [blue] orbitals):
 an $A_{1g}$ mode can be excited from them through a first-order Raman
 process.
 In real space, this is represented by the fact that the color of the
 points does not change while emitting an $A_{1g}$ mode.
 }
 \label{fig:bzh}
\end{figure}
%%%%%%%%%%%%%%%%%%%%%%%%%%%%%

Suppose that an electron has been excited into the
$\pi^*$-band by a laser light [Fig.~\ref{fig:bzh}(a)].
When a photo-excited electron emits an $A_{1g}$ mode,
there is a strong probability that 
the electron will undergo (intervalley) backward scattering,
as shown by the real space diagram in Fig.~\ref{fig:bzh}(b).
In the $k$-space,
the change in the exact (approximate) intervalley backward scattering
is denoted by the black (orange and green) solid arrow.
Although the forward scattering denoted by the dashed arrow 
may be allowed by momentum conservation,
it never takes place because orbitals suppress
the corresponding electron-phonon matrix element.
This dominance of intervalley backward scattering
originates from the characteristic feature of an $A_{1g}$ mode, namely
that the vibration consists only of bond shrinking/stretching motions, 
as shown by the displacement vectors in Fig.~\ref{fig:unit}.
Mathematically, this characteristic of the $A_{1g}$ mode
is described by the fact that the electron-phonon interaction, 
$\hat{H}_{\rm ep}(D)$, satisfies
\begin{align}
  & \chi_{\rm A}^\dagger\hat{H}_{\rm ep}(D)\chi_{\rm B}=
  \chi_{\rm B}^\dagger\hat{H}_{\rm ep}(D)\chi_{\rm A} =g_{\rm ep}, \\
  & \chi_{\rm A}^\dagger\hat{H}_{\rm ep}(D)\chi_{\rm A}=
\chi_{\rm B}^\dagger\hat{H}_{\rm ep}(D)\chi_{\rm B}=0,
\end{align}
where 
$g_{\rm ep}$ is a coupling constant for bond stretching.
With these $\hat{H}_{\rm ep}(D)$ conditions,
we obtain the electron-phonon matrix element squared 
$|M|^2=|\psi_{\rm K'}^{s'}(\Theta')^\dagger\hat{H}_{\rm ep}(D)\psi_{\rm
K}^s(\Theta)|^2$ using Eqs.~(\ref{eq:wf1}) and (\ref{eq:wf2}) as
\begin{align}
 |M|^2 = \frac{g^2_{\rm ep}}{2} \left\{
 1-ss' \cos(\Theta'-\Theta)\right\}.
 \label{eq:mateleD}
\end{align}
Equation~(\ref{eq:mateleD}) shows that, 
for intraband scattering ($ss'=1$),
the scattering probability of the exact backward scattering
($\Theta'=\Theta+\pi$) is maximum, while that of the exact forward
scattering ($\Theta'=\Theta$) vanishes.

When the energy of a photo-excited electron 
is much larger than the phonon energy ($\simeq$ 0.15 eV),
we can assume that 
no significant energy shift occurs as a result of the inelastic scattering. 
For the exact backward scattering,
the $A_{1g}$ wave vector relates to the photo-excited electron
wave vector as ${\bf q}=-2{\bf k}$,
where ${\bf q}$ (${\bf k}$) is the wave vector of the $A_{1g}$ mode
(photo-excited electron) measured from the K point.
This relationship between ${\bf k}$ and ${\bf q}$
shows that the orbitals effectively relate the electron wave vector ${\bf k}$
to the $A_{1g}$ wave vector ${\bf q}$.\footnote{Thus, the orbital
dependence of the electron-phonon matrix element justifies the basic
idea of the quasi-selection rule for the $D$ band proposed by several
authors~\cite{baranov87,pocsik98}.
The relationship between ${\bf k}$ and ${\bf q}$ ( ${\bf q}=-2{\bf k}$)
shows that $\hbar v_{\rm F}|{\bf q}|$ changes linearly
with changing excitation energy $E_L$ because 
$2\hbar v_{\rm F}|{\bf k}|$ is approximately equal to $E_L$.
It becomes important when we discuss the dispersive behavior of the $D$
band in Sec.~\ref{ssec:db}.
}

It is straightforward to reproduce Eq.~(\ref{eq:mateleD})
in the framework of the graphene Dirac equation.
The electron-phonon interaction for an $A_{1g}$ mode with wave vector
${\bf q}$ is written as~\cite{sasaki08ptps}
\begin{align}
 \hat{H}_{\rm ep}(D_{\bf q}) = g_{\rm ep}
 \begin{pmatrix}
  0 & e^{-i {\bf q}\cdot {\bf r}} \sigma_x \cr
  e^{i {\bf q}\cdot {\bf r}} \sigma_x & 0 
 \end{pmatrix},
 \label{eq:HDeff}
\end{align}
and the matrix element $M$
is given by
\begin{align}
 M 
 &=\int_V d^2{\bf r}
 \psi^{s'}_{\rm K'}({\bf r})^\dagger \hat{H}_{\rm
ep}(D_{\bf q}) \psi^s_{\rm K}({\bf r}) \nn \\
 &= \left(\frac{1}{V}
 \int_V d^2{\bf r} e^{i({\bf k+q-k'})\cdot {\bf r}}
 \right) \left(
 \psi_{\rm K'}^{s'}({\bf k'})^\dagger  \sigma_x \psi_{\rm K}^s({\bf k})
 \right).
\end{align}
The last line is written as a multiplication of two parts:
the first part represents momentum conservation
and the wave vector of the scattered electron ${\bf k'}$
is given by ${\bf k'}={\bf k}+{\bf q}$.
The second part gives Eq.~(\ref{eq:mateleD}).
In addition to momentum conservation,
we can use energy conservation to obtain 
$v_{\rm F}|{\bf k'}|=v_{\rm F}|{\bf k}|-\omega_{\bf q}$,
where $\omega_{\bf q}$ is the frequency of $A_{1g}$.
Thus, when $\hbar v_{\rm F}|{\bf k}| \gg \hbar \omega_{\bf q}$, 
we have $|{\bf k'}|\simeq |{\bf k}|$.
Since the orbital part results in the dominance of intervalley backward
scattering, we obtain ${\bf k'}\simeq {\bf -k}$. 
As a result, ${\bf q}\simeq -2{\bf k}$ is satisfied.

\subsection{Brillouin zone folding}\label{ssec:bzh}

The dominance of intervalley backward scattering 
causes an enhancement of the $D$ band Raman intensity 
if Brillouin zone folding (BZF) by the armchair edge is taken into
account~\cite{sasaki11_Dband}.
Here, BZF means that, in the BZ of graphene shown in Fig.~\ref{fig:unit},
a state with $(k_x,k_y)$ is identical to a state with $(-k_x,k_y)$
and that the correct BZ is given by 
the positive $k_x$ region of the original BZ of
graphene~\cite{sasaki11-armwf}.
In Fig.~\ref{fig:bzh}(c),
as a consequence of BZF, 
the final state in the intervalley backward scattering event 
is identified with the state near the K point.
Namely, the state with $(-|k_x|,-k_y)$
near the K$'$ point is identified with the state with
$(|k_x|,-k_y)$ near the K point.
So, in the folded BZ,
the change of a photo-excited electron is $(|k_x|,k_y) \to
(|k_x|,-k_y)$, as shown by the solid arrow in Fig.~\ref{fig:bzh}(c). 
Generally, the probability of a process 
in the folded BZ is given by replacing $\Theta'$ with $\pi-\Theta'$ in
Eq.~(\ref{eq:mateleD}) as
\begin{align}
 |M_{\rm BZF}|^2 = \frac{g^2_{\rm ep}}{2} \left\{
 1+ss' \cos(\Theta'+\Theta)\right\},
 \label{eq:DBZF}
\end{align}
and this transition probability is indeed maximum 
when $\Theta'=-\Theta$. 
When $\Theta'=\Theta$ and $s'=s$ in Eq.~(\ref{eq:DBZF}), 
the final state coincides with the initial state,
which is the condition of a first-order Raman process.
In this case, the probability is given by
\begin{align}
 |M_{\rm BZF}|^2 =g^2_{\rm ep}\cos^2\Theta,
 \label{eq:facBZF}
\end{align}
which takes its maximum value for $\Theta=0$ and $\pi$.
This shows that the states near the $k_x$-axis 
(or the states with bonding and antibonding orbitals)
can contribute to the $D$ band intensity through a first-order Raman process.

\begin{figure}[htbp]
 \begin{center}
  \includegraphics[scale=1.0]{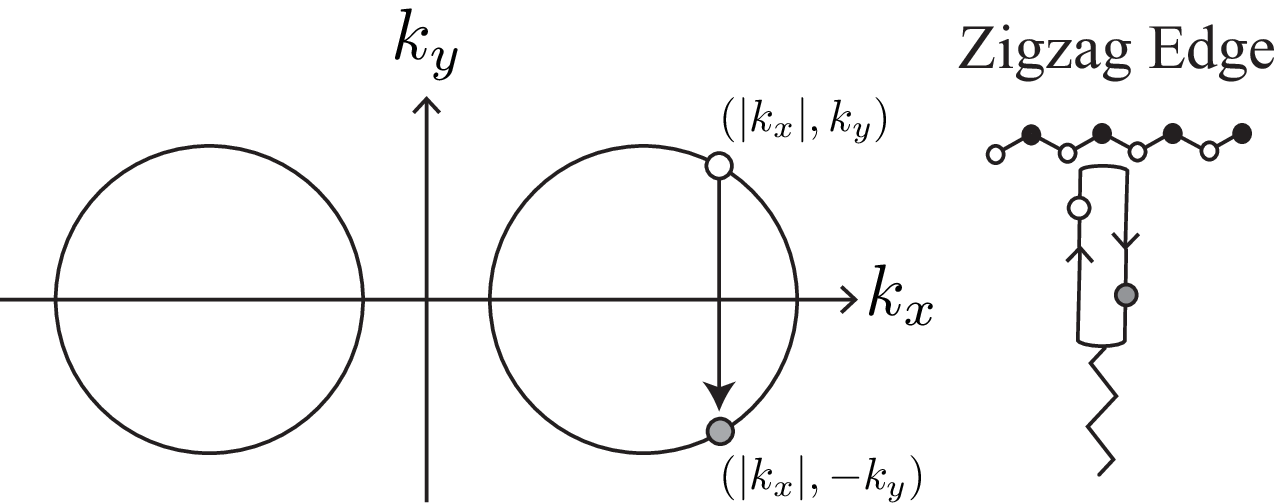}
 \end{center}
 \caption{A process that causes the $D'$ band at the zigzag edge.
 Phonon excitation (${\bf q}\ne 0$) accompanied by a change in
 the electronic states, which shows that the phonon does not appear
 through a first-order process.
 Note that, in the right diagram, the color of the points changes after
 the phonon is emitted.
 }
 \label{fig:zig}
\end{figure}

BZF originates from the fact that 
a special standing wave is formed by an armchair edge. 
The standing wave is constructed by the
antisymmetric combination of 
an incident plane wave with ${\bf k}=(k_x,k_y)$ 
and a scattered plane wave with ${\bf k}'=(-k_x,k_y)$ as
$e^{ik_y y}(e^{ik_x x}-e^{-ik_x x})\propto e^{ik_yy}\sin(k_x x)$.
A symmetric combination does not satisfy the boundary condition for the
armchair edge~\cite{sasaki11-armwf}.
Note that $\sin(k_x x)$ does not change when $k_x$ is replaced with $-k_x$,
except for the unimportant change in the overall sign.
More importantly, 
the orbital part Eq.~(\ref{eq:wf1}) does not change
when there is the reflection at the armchair edge, 
as we can confirm by replacing $\Theta'$ with
$\pi-\Theta$ in Eq.~(\ref{eq:wf2})~\cite{sasaki10-jpsj,sasaki11-armwf}.
The same orbitals are superposed to form a standing wave at the armchair
edge.
Thus, the total wave function becomes
\begin{align}
 \psi^s_{\rm a}({\bf k})
 =N \left\{
 e^{i{\bf k}\cdot {\bf r}} 
 \begin{pmatrix}
  e^{-i\Theta({\bf k})} \cr s
 \end{pmatrix}
 -
 e^{i{\bf k}'\cdot {\bf r}} 
 \begin{pmatrix}
  -e^{i\Theta'({\bf k}')} \cr s
 \end{pmatrix}
 \right\}
 =N'
 e^{ik_y y} \sin(k_x x) 
 \begin{pmatrix}
  e^{-i\Theta({\bf k})} \cr s
 \end{pmatrix},
 \label{eq:standA}
\end{align}
where $N$ ($N'$) is a normalization constant.
The standing wave does not change with the replacement 
$k_x \to -k_x$, and therefore the correct BZ of the standing wave is
given solely by the positive $k_x$ region to avoid double counting.
It is noteworthy that BZF is specific to the armchair edge and 
is not applicable to a zigzag edge.
The absence of BZF at a zigzag edge is due to the orbital part
changes with the reflection of an electron at the zigzag edge
and the different orbitals are superposed to form a standing wave~\cite{sasaki10-jpsj,sasaki10-forward}.
The standing wave for a zigzag edge is written as 
\begin{align}
 \psi^s_{\rm z}({\bf k})
 =N e^{ik_x x}\left\{
 e^{ik_y y} 
 \begin{pmatrix}
  e^{-i\Theta({\bf k})} \cr s
 \end{pmatrix}
 -
 e^{-ik_y y} 
 \begin{pmatrix}
  e^{i\Theta({\bf k})} \cr s
 \end{pmatrix}
 \right\}
 =N' 
 e^{ik_x x}
 \begin{pmatrix}
  \sin(k_y y -\Theta({\bf k}))) \cr s \sin(k_y y)
 \end{pmatrix},
\end{align}
which is not invariant with the replacement $k_y \to -k_y$ 
unless $\Theta= 0$ or $\pi$.
Thus, it needs a second-order process to activate a phonon
mode with nonzero ${\bf q}$ (see Fig.~\ref{fig:zig}) and 
a first-order Raman band (except the $G$ band) cannot 
appear at the zigzag edge.
Although the intensity is not comparable to that of $G$ or $D$ bands,
we can expect a second-order band to appear at the zigzag edge.
The $D'$ band~\cite{maeta77,nemanich79} (not the $D$ band)
may be such a second-order Raman band
that can be described by the double resonance model.

%%%%%%%%%%%%%%%%%%%%%%%%%%%%%
%%%%%%%%%%%%%%%%%%%%%%%%%%%%%

The standing wave at the armchair edge 
can be expressed in the framework of the Dirac equation
as
\begin{align}
 \psi^s_{{\rm a},{\bf k}}({\bf r}) 
 =N e^{ik_y y}
 \begin{pmatrix}
   e^{+ik_x x}e^{-i\Theta({\bf k})} \cr  e^{+ik_x x}s \cr e^{-ik_x x}e^{-i\Theta({\bf k})} \cr e^{-ik_x x}s 
 \end{pmatrix}
 = N e^{ik_y y}
 \begin{pmatrix}
  e^{+ik_x x} \cr e^{-ik_x x}
 \end{pmatrix}\otimes \frac{1}{\sqrt{2}}
 \begin{pmatrix}
  e^{-i\Theta({\bf k})} \cr s 
 \end{pmatrix},
 \label{eq:armDirac}
\end{align}
where ${\bf k}$ denotes the wave vector measured from the Dirac point in
the folded BZ and $\otimes$ represents the direct product of the valley
and the orbital.
It can be confirmed that 
Eq.~(\ref{eq:armDirac}) reproduces Eq.~(\ref{eq:DBZF}), 
by calculating the expectation value of $\hat{H}_{\rm ep}(D_{\bf q})$
of Eq.~(\ref{eq:HDeff}) with respect to Eq.~(\ref{eq:armDirac}),
\begin{align}
 M_{\rm BZF}
 &=\int_V d^2{\bf r}
 \psi^{s'}_{{\rm a},{\bf k'}}({\bf r})^\dagger \hat{H}_{\rm
ep}(D_{\bf q}) \psi^s_{{\rm a},{\bf k}}({\bf r}) \nn \\
 &= \delta(q_x-k_x-k'_x) \frac{1}{2}
 \left\{
 \delta(k_y +q_y-k'_y) 
 +\delta(k_y -q_y-k'_y) \right\}
 \frac{g_{\rm ep}}{2}\left(
 e^{i\Theta'}s + s'e^{-i\Theta}
 \right),
\end{align}
where we used $N^2\int dy e^{i(k_y\pm q_y -k'_y)y} \int dx e^{\pm
i(q_x-k_x-k'_x)x} =(1/2)\delta(k_y \pm q_y-k'_y)\delta(q_x-k_x-k'_x)$.
For ${\bf k'}={\bf k}$ (i.e., for a first order Raman process),
Eq.~(\ref{eq:facBZF}) is reproduced, and momentum conservation gives ${\bf q}=(2k_x,0)$.

%It is worth noting that Eq.~(\ref{eq:armDirac}) with $\Theta=0$ and
%$\pi$ are eigenstate of $\hat{H}_{\rm ep}(D(-2k_x,0))$.
%In adiabatic approximation,
%the bonding and antibonding orbitals of the standing wave functions
%are the eigenstates of the total Hamiltonian 
%$\hat{H}+\hat{H}_{\rm ep}(D(-2k_x,0))$, which is the most remakable
%property of the $A_{1g}$ mode.
%This is more than coincidence, as we have shown that in a previous
%paper, the presence of armchair edge can be modeled as a singlar
%mass term~\cite{sasaki10-chiral}.
%Although Eq.~(\ref{eq:armDirac}) looks reasonable, 
%we need to pay some special care when we use it,
%because Eq.~(\ref{eq:armDirac}) does not represent the sine
%function shown in Eq.~(\ref{eq:standA}).
%In scanning tunneling microscopy (STM) images,
%the presence of armchair edge 
%may be confirmed by the presence of line nodes.

\subsection{Optical anisotropy}\label{ssec:opan}

\begin{figure}[htbp]
 \begin{center}
  \includegraphics[scale=0.5]{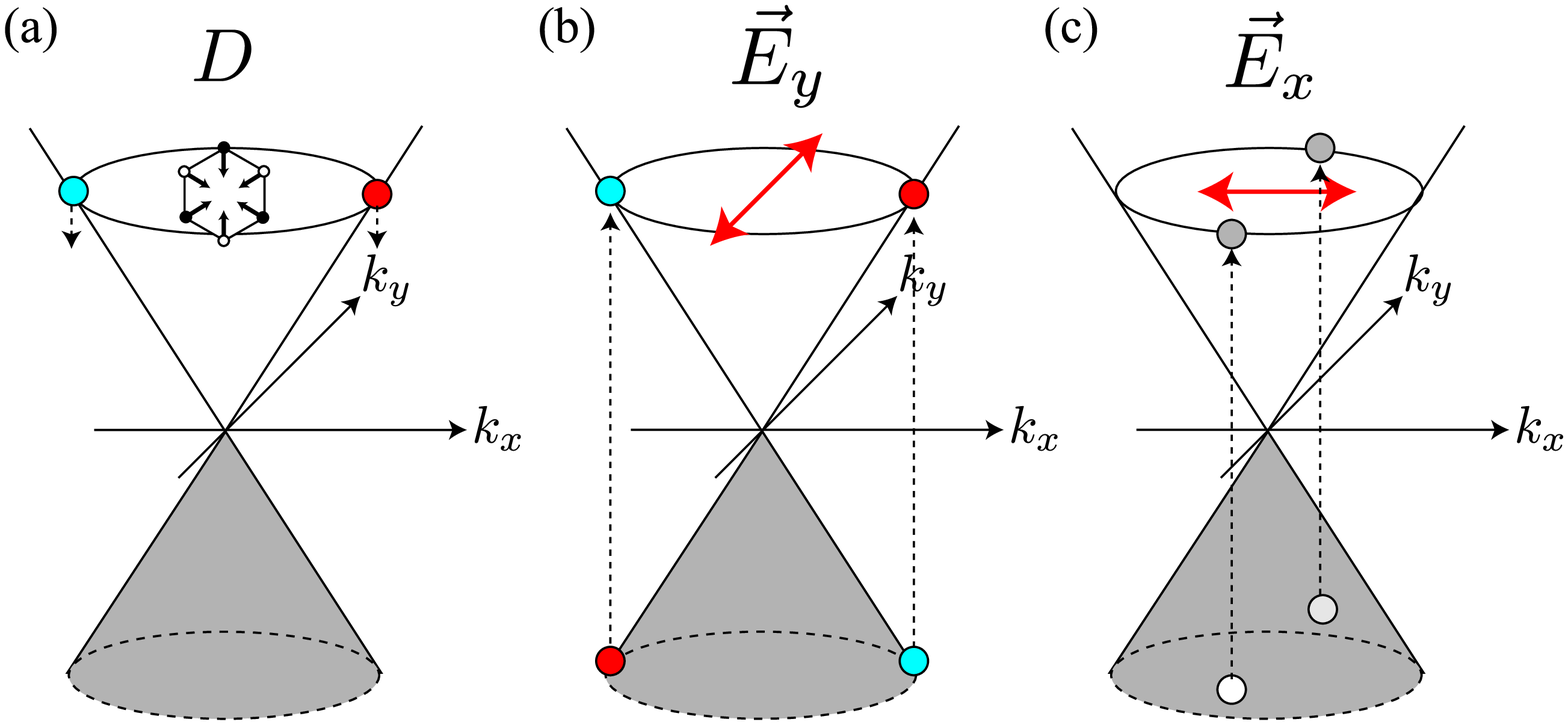}
 \end{center}
 \caption{Light polarization dependence of the $D$ band.
 (a) An $A_{1g}$ phonon is excited through a first-order process 
 from the electrons with bonding and antibonding orbitals on the $k_x$-axis.
 (b) To supply a photo-excited electron to the states on the
 $k_x$ axis, the laser light polarization must be parallel to the
 $y$-axis or the armchair edge.
 (c) $\vec{E}_x$ can produce a photo-excited electron at the $k_y$-axis.
 Such an electron changes its position when it emits an $A_{1g}$ mode,
 and thus it cannot contribute to the $D$ band.
 }
 \label{fig:pol}
\end{figure}

The third factor is the polarization dependence of optical transitions.
To supply photo-excited electrons
to the states with bonding and antibonding orbitals on the $k_x$-axis,
the polarization of incident laser light must be set 
parallel to the armchair edge ($y$-axis), as shown in
Fig.~\ref{fig:pol}(b).
The photo-excited electrons on the $k_x$-axis 
emit $A_{1g}$ modes without changing their positions [see Fig.~\ref{fig:pol}(a)].
Meanwhile, when the polarization of the incident laser light is set 
perpendicular to the armchair edge, the $x$-polarized light
supplies the states on the $k_y$-axis with 
photo-excited electrons [see Fig.~\ref{fig:pol}(c)].
The electrons near the $k_y$-axis change their positions in
the folded BZ when they emit $A_{1g}$ modes [see Fig.~\ref{fig:bzh}(b)],
and these electrons on the $k_y$-axis do not contribute to the $D$ band
intensity.
As a result, 
the $D$ band can be strongly dependent on the laser light polarization:
the $D$ band intensity is enhanced (suppressed) 
when the polarization of the incident laser light is parallel (perpendicular)
to the armchair edge.

The optical anisotropy is due to the $\Theta$ dependence of the optical
matrix element~\cite{grueneis03}.
Namely, for $y$ ($x$) polarized light, $\vec{E}_y$ ($\vec{E}_x$), 
the optical matrix element squared is proportional to $\cos^2\Theta$
($\sin^2\Theta$) as
\begin{align}
 |M_{\rm opt}|^2 =
\begin{cases}
 & g_{\rm e\gamma}^2 \cos^2\Theta \ \ \ {\rm for} \  \vec{E}_y, \\
 & g_{\rm e\gamma}^2 \sin^2\Theta \ \ \ {\rm for} \  \vec{E}_x,
\end{cases}
 \label{eq:optpol}
\end{align}
where $g_{\rm e\gamma}$ denotes an electron-light coupling constant.
The electrons near the $k_x$ ($k_y$)-axis are
dominantly photo-excited by $\vec{E}_y$ ($\vec{E}_x$) [see
Fig.~\ref{fig:pol}(b) and (c)].
For the general polarization direction of incident laser light ${\bf E}$,
$M_{\rm opt}$ is proportional to the vector product of ${\bf E}$ and
$\hat{\bf k}$ ($\equiv {\bf k}/|{\bf k}|$) as 
$M_{\rm opt} \propto {\bf E} \times \hat{\bf k}$.

%%%%%%%%%%%%%%%%%%%%%%%%%%%%%
%%%%%%%%%%%%%%%%%%%%%%%%%%%%%

Although Eq.~(\ref{eq:optpol}) was derived without taking account of the edge,
it turns out that a similar optical matrix element is obtained for the standing
waves~\cite{sasaki11-dc}. 
Here, let us use the Dirac equation of Eq.~(\ref{eq:H0}) 
to obtain the matrix element that includes the effect of the armchair edge.
The electron-light interaction is given by replacing 
$\hat{\bf p}$ with $\hat{\bf p}-e{\bf A}$ in Eq.~(\ref{eq:H0}) as
\begin{align}
 \hat{H} =
 v_{\rm F} 
 \begin{pmatrix}
  \bsigma \cdot (\hat{\bf p}-e{\bf A}) & 0 \cr
  0 & \bsigma' \cdot (\hat{\bf p}-e{\bf A}) \cr
 \end{pmatrix},
 \label{eq:Hem}
\end{align}
where ${\bf A}$ is the vector potential of light.
Since the Maxwell equation gives ${\bf E} = - \partial{\bf A}/\partial t$,
the vector directions of ${\bf E}$ and ${\bf A}$ are the same.
The optical matrix element is defined
using Eq.~(\ref{eq:armDirac}) as
\begin{align}
 M_{\rm opt} = -e  v_{\rm F} 
 \int d^2{\bf r} \psi_{{\rm a},{\bf k'}}^{+1}({\bf r})^\dagger 
 \begin{pmatrix}
  \bsigma \cdot {\bf A} & 0 \cr
  0 & \bsigma' \cdot {\bf A} \cr
 \end{pmatrix}
 \psi_{{\rm a},{\bf k}}^{-1}({\bf r}).
\end{align}
For the $y$-polarized light ${\bf A}=(0,A_y)$, $M_{\rm opt}$
is nonzero only for a direct transition (${\bf k'}={\bf k}$) and the
orbital gives $\cos\Theta$.
For the $x$-polarized light ${\bf A}=(A_x,0)$, $M_{\rm opt}$
includes the integral $(1/L)\int_0^L dx \sin (k_x -k'_x)x $ and 
the orbital gives $(-e^{i\Theta'} + e^{-i\Theta})/2$.
Since the integral vanishes when $k_x=k'_x$, 
a direct transition does not take place.
The possible transitions are indirect transitions $k_x\ne k'_x$.
For $k_x-k'_x=n\pi/L$ ($n$ is an odd number),
we obtain $2/(n \pi)$ by performing the integral.
The momentum change in an indirect transition is 
inversely proportional to the distance ($L$) from the armchair edge 
where the standing wave is a good approximation.
Since the change in the wave vector is negligible when $L$ is large,
we may assume $\Theta' \simeq \Theta$.
Then the orbital leads to $-i\sin \Theta$, which reproduces Eq.~(\ref{eq:optpol}).
This feature of the indirect transition for the $x$-polarized light
has been examined with a more mathematically rigorous method using a
lattice tight-binding model~\cite{sasaki11-dc}.

\subsection{$D$ band polarization formula}\label{ssec:fin}

\begin{figure}[htbp]
 \begin{center}
  \includegraphics[scale=0.6]{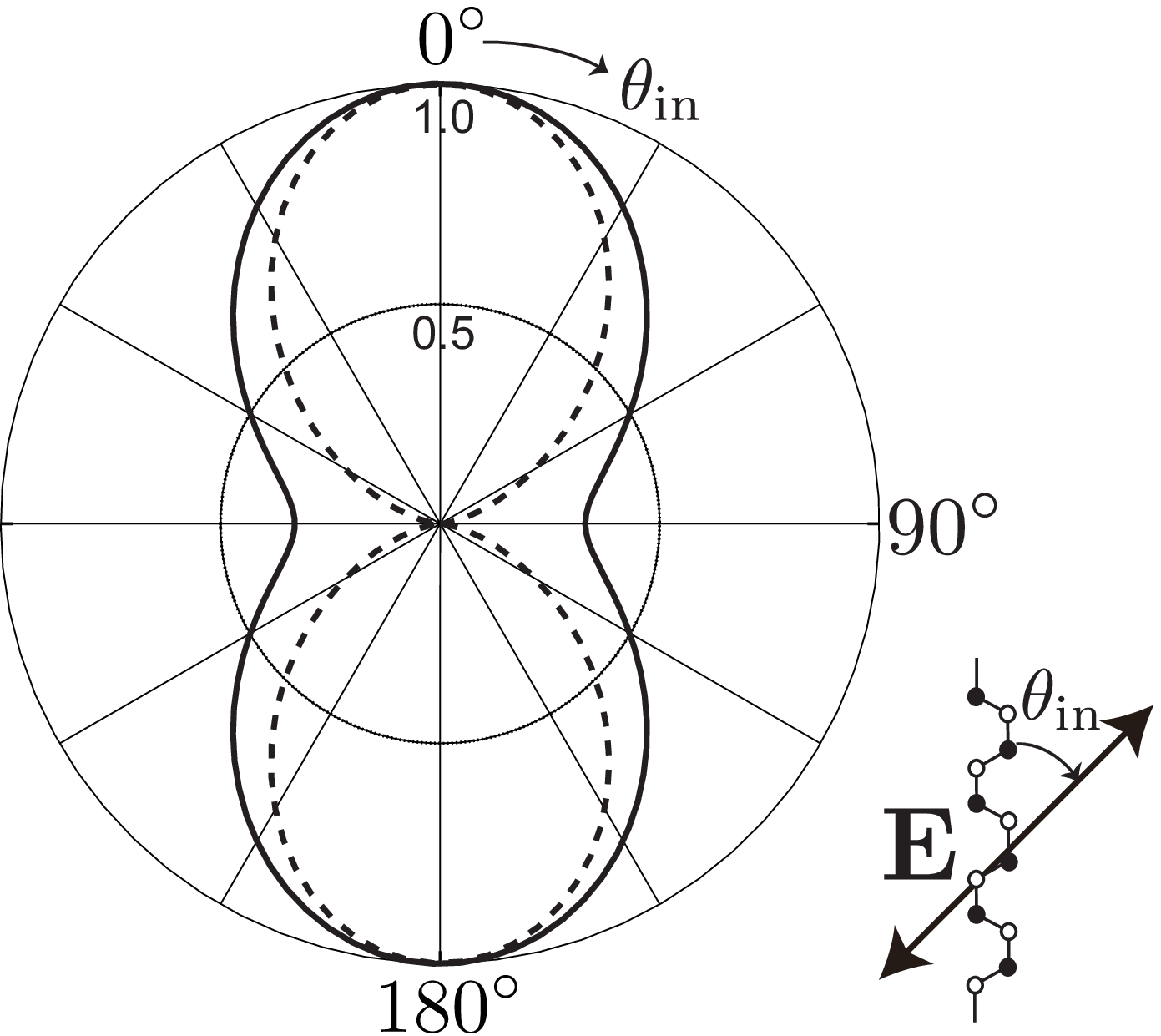}
 \end{center}
 \caption{Polar plot for the $D$ band intensity.
 The parameters for the solid curve are $a=1$, $b=1/3$, and $c=0$,
 while those for the dashed curve are $a=1$ and $b=c=0$.
 }
 \label{fig:polarp}
\end{figure}

We combine Eqs.~(\ref{eq:facBZF}) and (\ref{eq:optpol})
to derive the polarization formula of the $D$ band.
The probability of a first-order Raman process that 
an electron with $\Theta$ in the $\pi$-band
is excited into the $\pi^*$-band by $\vec{E}_i$ ($i=\{x,y\}$), 
and then the photo-excited electron emits the $A_{1g}$ mode, 
and finally the electron with $\Theta$ emits a light with $\vec{E}_j$ is
given by 
\begin{align}
 |M_{ji}(\Theta)|^2 = |M_{\rm opt}(\vec{E}_j)|^2 |M_{\rm BZF}(\Theta)|^2  |M_{\rm opt}(\vec{E}_i)|^2 
 =g_{\rm e\gamma}^4 g^2_{\rm ep}
 \begin{cases}
  \cos^6\Theta \ \ \ &(ji)=(yy) \\
  \cos^4\Theta \sin^2\Theta \ \ \ &(xy)\ {\rm or}\ (yx) \\
  \cos^2\Theta \sin^4\Theta \ \ \ &(xx). \\
 \end{cases}
\end{align}
Note that the wave vector of the $A_{1g}$ mode is completely fixed by $\Theta$ 
and $|{\bf k}|$ (or $E_L$) as 
\begin{align}
 q_x=-2|{\bf k}| \cos\Theta.
 \label{eq:phwavevec}
\end{align}
Phonons with different momenta are distinguishable in principle.
We have different final states for different $\Theta$ values, and 
to calculate the $D$ band intensity,
we need to sum over all possible final states by operating 
$|M_{ji}(\Theta)|^2$ with $\int_0^{2\pi} d\Theta$.
Because $\int_0^{2\pi} \cos^2 \Theta
\sin^4 \Theta d\Theta/\int_0^{2\pi} \cos^6 \Theta d\Theta=1/5$,
the polarization dependence of the $D$ band intensity is written as
\begin{align}
 I_D(\theta_{\rm out},\theta_{\rm in})
 \propto 
 \begin{pmatrix}
  \cos^2 \theta_{\rm out} & \sin^2 \theta_{\rm out} 
 \end{pmatrix}
 \begin{pmatrix}
  1 & 1/5 \cr 1/5 & 1/5
 \end{pmatrix}
 \begin{pmatrix}
  \cos^2 \theta_{\rm in} \cr \sin^2 \theta_{\rm in} 
 \end{pmatrix},
\end{align}
where $\theta_{\rm in}$ ($\theta_{\rm out}$) 
denotes the angle of an incident (scattered) electric field
with respect to the armchair edge.\footnote{
In calculating the $D$ band Raman intensity,
it is incorrect to sum over intermediate states
specified by the electron's wave vector ${\bf k}$,
such as $|\sum_{\bf k}M_{ji}(\Theta({\bf k}))|^2$.
Because the phonon wave vector relates to the electron wave vector
through ${\bf q}=-2{\bf k}$, $\sum_{\bf k}$ actually means 
a summation for different phonons (final states).
Thus, the Raman intensity is proportional to 
$\sum_{\bf q}|M_{ji}(\Theta({\bf q}))|^2$ or $\int_0^{2\pi} d\Theta |M_{ji}(\Theta)|^2$.
}$^{,}$\footnote{
The polarization dependence of the $D$ band near the edge has
been calculated by Basko~\cite{basko09_bound}, and the result is
different from our result.
The difference may arise from the fact that here we did not sum over
the intermediate states nor take into account the energy denominators
of the perturbation theory; instead, we have assumed a resonant Raman
process, that is, we select a particular intermediate state for each
final phonon state.
We give some notes concerning resonant condition in Appendix~\ref{app:rc}.
}
When the VV configuration ($\theta_{\rm out} = \theta_{\rm in}$) is used, 
the polarization dependence is approximated by
$I^{VV}_D(\theta_{\rm in}) \simeq I \cos^4 \theta_{\rm in}$.
On the other hand, when the VH configuration ($\theta_{\rm out} =
\theta_{\rm in} + \pi/2$) is used, 
the polarization dependence is approximated by
$I^{VH}_D(\theta_{\rm in}) \simeq I \cos^2 \theta_{\rm in}\sin^2
\theta_{\rm in}$.
These results are consistent with the experiment reported by
Can\ifmmode \mbox{\c{c}}\else \c{c}\fi{}ado et al~\cite{canifmmode04sec}.
Without a polarizer for the scattered light,
we have 
\begin{align}
 I_D(\theta_{\rm in}) \propto 
 \left( \int_0^{2\pi} d\Theta \cos^4\Theta \right)
 \cos^2(\theta_{\rm in})+ \left(
 \int_0^{2\pi} d\Theta \cos^2 \Theta \sin^2 \Theta \right)
 \sin^2(\theta_{\rm in}).
\end{align}
Because $\int_0^{2\pi} \cos^2 \Theta
\sin^2 \Theta d\Theta/\int_0^{2\pi} \cos^4 \Theta d\Theta=1/3$,
it may be rewritten as
$I_D(\theta_{\rm in}) \propto
\cos^2(\theta_{\rm in})+(1/3)\sin^2(\theta_{\rm in})$.
With (Without) a polarizer for scattered light, 
the depolarization ratio $I_D(90^\circ)/I_D(0^\circ)$ is $1/5$ ($1/3$).
Generally, the $D$ band polarization dependence
is fitted using an empirical formula 
\begin{align}
 I_D(\theta_{\rm in}) \propto
 \cos^2(\theta_{\rm in})+b\sin^2(\theta_{\rm in})+c,
 \label{eq:geneformu}
\end{align}
where $c$ is a constant, which probably originates from a defect beside
the edge. 
The constants $b$ and $c$ determine the depolarization ratio:
$I_D(90^\circ)/I_D(0^\circ)=(b+c)/(1+c)$ [see Fig.~\ref{fig:polarp}].
When $c$ is much larger than unity, the polarization behavior is obscured.
If $c$ is negligible, 
a smaller depolarization ratio ($b=1/5$) is expected 
when a polarizer is used for the scattered light.

%%%%%%%%%%%%%%%%%%%%%%%%%%%%%
%%%%%%%%%%%%%%%%%%%%%%%%%%%%%

\section{Predictions of our model}\label{sec:pre}

We have seen that 
the $D$ band has a direct relationship to the bonding and antibonding
orbitals (that is, an $A_{1g}$ mode is excited
through the first-order Raman process only from these orbitals). 
This conclusion has been derived based on two factors: 
the dominance of intervalley backward scattering (Sec.~\ref{ssec:back}),
and BZF by the armchair edge (Sec.~\ref{ssec:bzh}).
In this section, 
we report some consequences that are derived from these factors.

\subsection{The origin of dispersive behavior}\label{ssec:db}

The $D$ band frequency $\omega_D$ increases linearly with increasing
excitation energy $E_L$ as $\partial \omega_D/\partial E_L \sim 50$
cm$^{-1}$/eV, which is known as dispersive behavior~\cite{vidano81,matthews99,gupta09,casiraghi09}.
In a previous paper~\cite{sasaki12_migration}, 
we pointed out that the dispersive behavior is mainly attributed to 
a quantum mechanical correction (self-energy)
to the $A_{1g}$ frequency.
The modified energy of the $A_{1g}$ mode is written as
$\hbar \omega + {\rm Re} \Pi_\mu(q,\omega)$, where $\omega$ is the bare
frequency\footnote{
Calculation suggests that the bare frequency contains a term quadratic
in $q$, which is consistent with inelastic x-ray scattering data for graphite~\cite{grueneis09}.}  
and $\Pi_\mu(q,\omega)$ represents the self-energies of the
$A_{1g}$ modes induced by electron-phonon interaction.
The self-energy of an $A_{1g}$ mode with $q=|{\bf q}|$ is defined as
\begin{align}
 \Pi_\mu(q,\omega) 
 \equiv 
 \sum_{s,s'} \sum_{\bf k} 
 \frac{(f^{s}_{{\bf k},\mu}-f^{s'}_{{\bf k+q},\mu}) g_s |M|^2}{\hbar\omega+s\hbar v_{\rm F}k-s'\hbar
 v_{\rm F}|{\bf k+q}|+i\epsilon},
 \label{app:Pi}
\end{align}
where 
$\epsilon$ is a positive infinitesimal,
$f^s_{{\bf k},\mu}=\lim_{\beta\to\infty}
(1+e^{\beta (s v|{\bf k}|-\mu)})^{-1}$ 
is the Fermi distribution function
with finite doping $\mu$,
and $g_{s}=2$ represents spin degeneracy.
The electron-phonon matrix element squared 
$|M|^2$ is constructed in Eq.~(\ref{eq:mateleD})
as 
$|M|^2 = \frac{g_{\rm ep}^2}{2}
 \left\{
 1-ss'\cos(\Theta'({\bf k+q})-\Theta({\bf k})) \right\}$.
%We can assume $\mu \ge 0$ without losing generality 
%because of particle-hole symmetry.
%defined at zero temperature 

%%%%%%%%%%%%%%%%%%%%%%%%%%%%%
\begin{figure}[htbp]
 \begin{center}
  \includegraphics[scale=0.6]{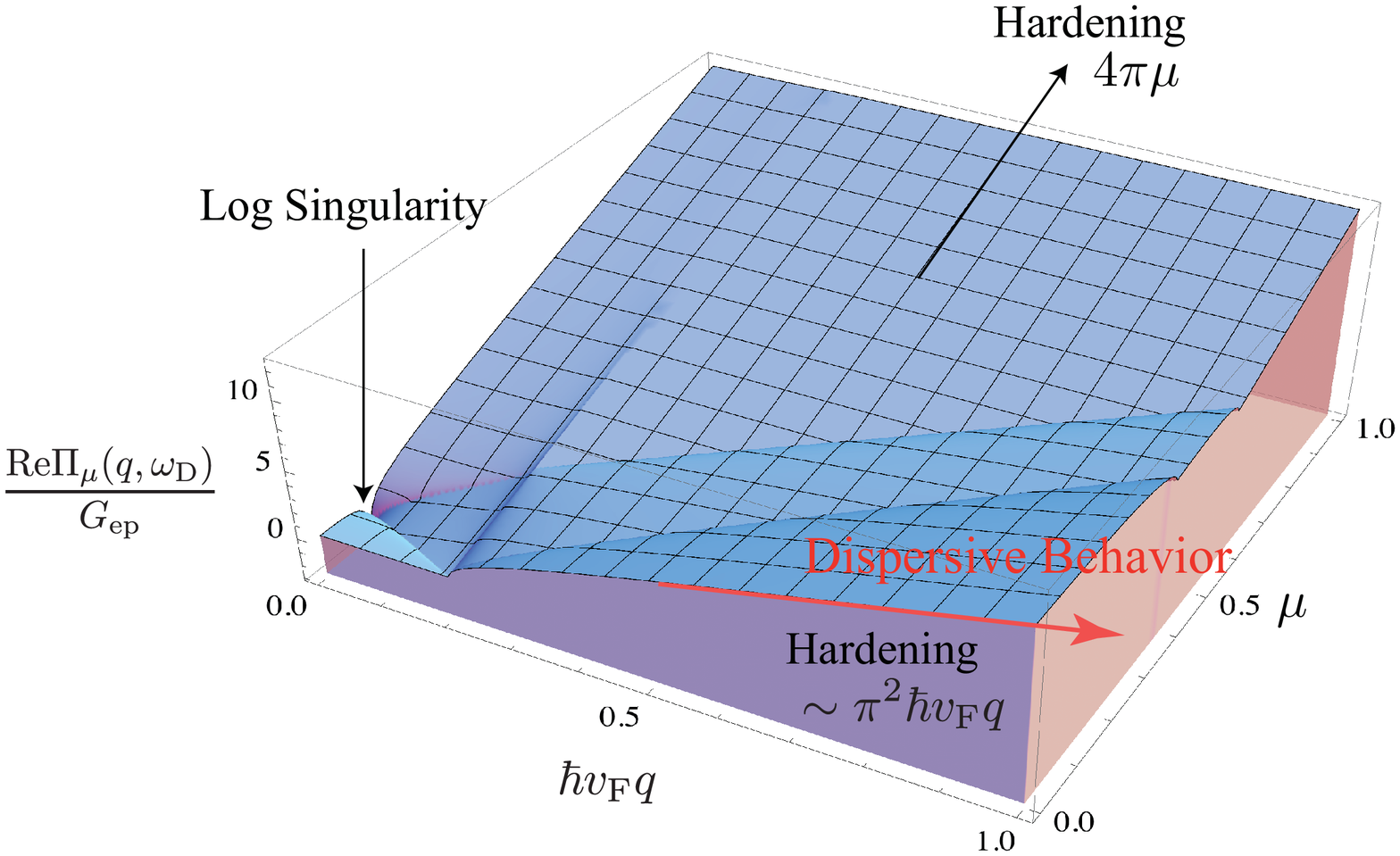}
 \end{center}
 \caption{
 A 3d plot of ${\rm Re}\Pi_\mu(q,\omega_D)$.
 The variables $\hbar v_{\rm F}q$ and $\mu$ are given in eV. 
 Note that ${\rm Re}\Pi_\mu(q,\omega)$ does not include the $q$
 dependence of the bare frequency.
 }
 \label{fig:rePI}
\end{figure}
%%%%%%%%%%%%%%%%%%%%%%%%%%%%%

In the continuum limit of ${\bf k}$, 
$\Pi_\mu(q,\omega)$ is calculated analytically.
The expression of the real part is 
given by
\begin{align}
 &{\rm Re}\Pi_\mu(q,\omega)/G_{\rm ep} = 4 \pi \mu  \nn \\
 & \ +\hbar \pi \sqrt{\omega^2-v^2_{\rm F}q^2}\theta_{\omega-v_{\rm F}q}
 \left[-g\left(\frac{\hbar\omega+2\mu}{\hbar v_{\rm F}q}\right)
 +\theta_{\frac{\hbar\omega-\hbar v_{\rm F}q}{2}-\mu}
 g\left(\frac{\hbar\omega-2\mu}{\hbar v_{\rm F}q}\right)
 +\theta_{\mu-\frac{\hbar\omega+\hbar v_{\rm F}q}{2}}g\left(\frac{2\mu
 -\hbar\omega}{\hbar v_{\rm F}q}\right)
 \right] \nn \\
 & \  
 + \hbar\pi \sqrt{v_{\rm F}^2q^2-\omega^2}\theta_{v_{\rm F}q-\omega} \left\{
 \theta_{\frac{\hbar v_{\rm F}q-\hbar \omega}{2}-\mu}
 \left[\frac{\pi}{2} - \sin^{-1}\left(\frac{\hbar \omega+2\mu}{\hbar v_{\rm F}q}\right)\right] 
 + \theta_{\frac{\hbar v_{\rm F}q+\hbar\omega}{2}-\mu}\left[
 \frac{\pi}{2}-\sin^{-1}\left(\frac{2\mu-\hbar\omega}{\hbar v_{\rm F}q}
 \right)\right] \right\},
\label{eq:rePI}
\end{align}
where $\theta_x$ denotes a step function satisfying 
$\theta_{x\ge 0}=1$ and $\theta_{x<0}=0$,
$g(x)\equiv \ln(x+\sqrt{x^2-1})$, and 
$G_{\rm ep} \equiv g_{\rm ep}^2V/(2\pi \hbar v_{\rm F})^2$ is a
(dimensionless) coupling constant.
A 3d plot of ${\rm Re}\Pi_\mu(q,\omega_D)/G_{\rm ep}$
is shown in Fig.~\ref{fig:rePI}.
Interestingly, ${\rm Re}\Pi_\mu(q,\omega_D)$ increases as we increase $q$.
In fact, near the Dirac point, ${\rm Re}\Pi_\mu(q,\omega_D)$ follows
\begin{align}
 {\rm Re} \Pi_{\mu\sim 0}(q,\omega) \simeq G_{\rm ep} \pi^2 \hbar v_{\rm F} q.
\end{align}
Since $\hbar v_{\rm F}q \simeq E_L$, 
the self-energy contributes to the dispersive behavior of the $D$ (or
$2D$) band~\cite{vidano81,matthews99,piscanec04,gupta09,casiraghi09}. 
If we use $G_{\rm ep}=5$ cm$^{-1}/{\rm eV}$, which is obtained from 
the broadening data published by Chen et al.~\cite{chen11},
the self-energy can account for $\sim$60\% of the dispersion
because ${\rm Re}\Pi_{\mu\simeq 0}(q,\omega_q)\simeq G_{\rm ep}\pi^2 E_L$
and $G_{\rm ep}\pi^2\simeq 30$ cm$^{-1}/{\rm eV}$.
It should be emphasized that
the self-energy is calculated using only Eq.~(\ref{eq:mateleD}), and any
artificial assumption, such as an adiabatic approximation,
is not employed when calculating the self-energy.
The physical origin of the dispersive behavior is easy to be understood
with shifted Dirac cones~\cite{sasaki12_migration}. 
In perturbation theory, the mechanism of dispersive behavior is almost
the same as the mechanism where the $G$ band exhibits hardening with increasing $|\mu|$.

The dispersive behavior is not an inherent property of the $D$ band but
rather is a property of the $A_{1g}$ mode.
The armchair edge is involved in the activation of the $D$ band.
However, the dispersive behavior itself has nothing to do with the edge.
In other words, 
there are processes that can exhibit dispersive behavior, besides the Raman $D$ band.
A good example is the $2D$ band, for which dispersive behavior is
observed~\cite{vidano81} because the $2D$ band consists of two $A_{1g}$ modes.

\subsection{Intravalley phonons}

In addition to the zone-boundary $A_{1g}$ mode,
we can determine the effect of BZF on zone-center (intravalley) optical phonon modes:
BZF forbids an intravalley transverse optical (TO) mode 
to appear as a prominent Raman band at the armchair edge.
The Dirac equation is the most useful way of showing this.
The electron-phonon interactions for the LO and TO modes 
with (nonzero) momentum $(q_x,q_y)$ are written as
\begin{align}
 & \hat{H}_{\rm LO}({\bf q}) = g_{\rm ep} \frac{e^{i{\bf q}\cdot {\bf r}}}{|{\bf q}|}
 \begin{pmatrix}
  \sigma_x q_y - \sigma_y q_x & 0 \cr
  0 & \sigma_x q_y + \sigma_y q_x
 \end{pmatrix},
 \\
 & \hat{H}_{\rm TO}({\bf q}) = g_{\rm ep} \frac{e^{i{\bf q}\cdot {\bf r}}}{|{\bf q}|}
 \begin{pmatrix}
  \sigma_x q_x + \sigma_y q_y & 0 \cr
  0 & \sigma_x q_x - \sigma_y q_y
 \end{pmatrix},
\end{align}
and their matrix elements are obtained from 
Eq.~(\ref{eq:armDirac}) as
\begin{align}
 M_{\rm LO}({\bf q}) &= \frac{g_{\rm ep}}{|{\bf q}|}
 \delta(-k'_y+q_y +k_y) \times \nn \\
 & \left\{ 
 \delta(-k'_x+q_x +k_x) \left[ \langle \sigma_x \rangle q_y -
 \langle \sigma_y \rangle q_x \right]
 +\delta(-k'_x-q_x +k_x) \left[ \langle \sigma_x \rangle q_y + 
 \langle \sigma_y \rangle q_x \right]
 \right\},
 \\
 M_{\rm TO}({\bf q}) &= \frac{g_{\rm ep}}{|{\bf q}|}
 \delta(-k'_y+q_y +k_y) \times \nn \\
 & \left\{ 
 \delta(-k'_x+q_x +k_x) \left[ \langle \sigma_x \rangle q_x + 
 \langle \sigma_y \rangle q_y \right]
 +\delta(-k'_x-q_x +k_x) \left[ \langle \sigma_x \rangle q_x - 
 \langle \sigma_y \rangle q_y \right]
 \right\},
\end{align}
where $\langle \sigma_x \rangle \equiv (se^{i\Theta'}+s' e^{-\Theta})/2$
and $\langle \sigma_y \rangle \equiv -i(se^{i\Theta'}-s' e^{-\Theta})/2$.
For the TO mode, 
$M_{\rm TO}({\bf q})$ vanishes in the $q_x \to 0$ limit,
due to the interference between the valleys.
In the $q_y \to 0$ limit, $\Theta'=\pi-\Theta$ holds, and 
the orbital of the matrix element, $\langle \sigma_x \rangle$,
vanishes when $s'=s$.
%This holds in the limit of $q_x\to 0$. 
Thus, the electron-phonon matrix element for the $\Gamma$ point TO mode 
is suppressed compared with that of the LO mode:
the $\Gamma$ point TO mode is missing in the $G$ band 
at the armchair edge~\cite{sasaki10-jpsj,cong10,begliarbekov10,zhang11}.
%Although the $D$ band is a first-order Raman band, 
%it can exhibit the dispersive behavior
%because the phonon wave vector $q_x$ corresponds to $2|k_x|$.
%This is contrast to that the phonon wave vector vanishes for the G band
%irrespective of the value of $k_x$, whereby the G band spectrum is
%insensitive to the laser excitation energies.

\subsection{$D$ band splitting}

\begin{figure}[htbp]
 \begin{center}
  \includegraphics[scale=0.8]{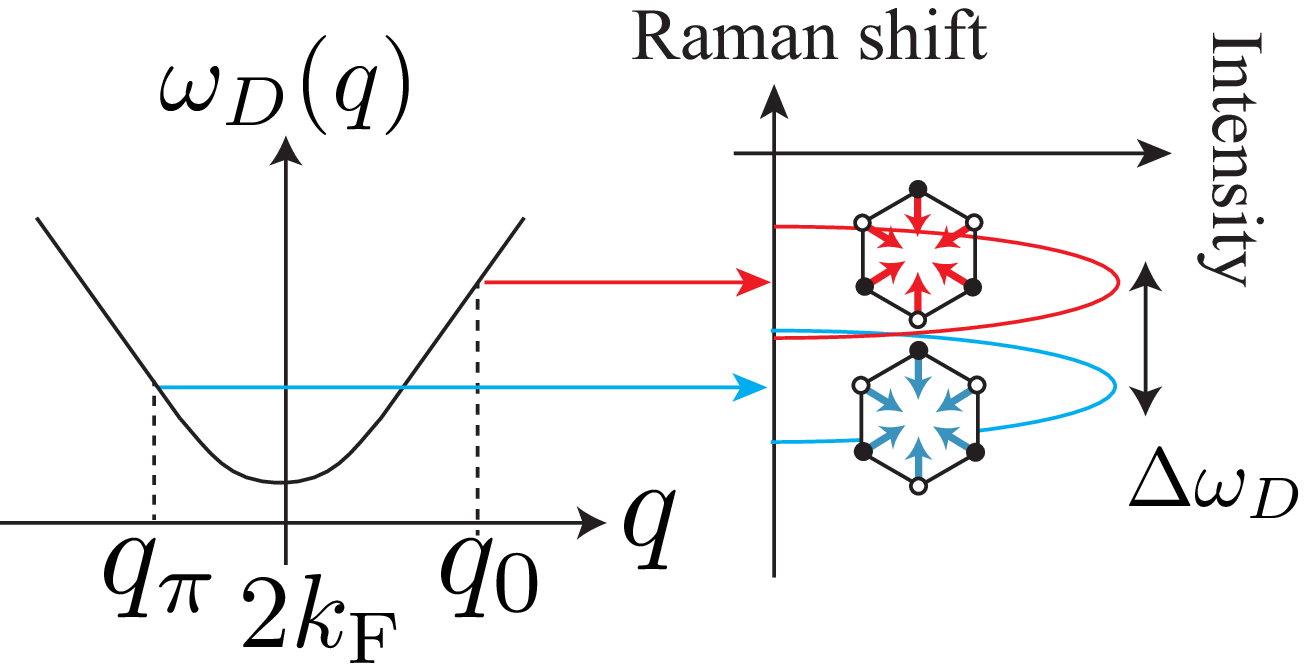}
 \end{center}
 \caption{
 The phonon dispersion relation near the Dirac point, where
 two principal wave vectors of the phonons ($q_0$ and $q_\pi$) contribute
 to the D band. 
 The energy difference between $\omega_D(q_0)$ and $\omega_D(q_\pi)$
 appears as two peaks in the $D$ band.
 }
 \label{fig:split}
\end{figure}

The $D$ band is composed of the (two) $A_{1g}$ modes
that are emitted from two electronic states with bonding or
antibonding orbitals ($\Theta=0$ or $\pi$).
This fact results in the splitting of the $D$ band 
if the trigonal warping effect is taken into account~\cite{sasaki11_Dband}.
The splitting width increases with
increasing incident laser energy $E_L$ as
\begin{align}
 \Delta \omega_D = 25 \left(\frac{E_L}{\gamma_0}\right)^2 \ 
 [{\rm cm}^{-1}],
 \label{eq:disp}
\end{align}
where $\gamma_0$ ($\simeq 3$ eV)
is the hopping integral between nearest neighbor atoms.
This formula is derived by noting 
that the wave vectors for the two $A_{1g}$ modes that originate from the
bonding and antibonding orbitals, are different due to the trigonal warping
effect.
The difference is estimated with 
the lattice tight-binding model as~\cite{sasaki11_Dband}
\begin{align}
 \Delta q = \frac{\sqrt{3}}{a} \left(\frac{E_L}{3\gamma_0}\right)^2.
 \label{eq:dtheta_p}
\end{align}
For simplicity, let us assume that 
the phonon dispersion relation for the $D$ band 
is isotropic about the Dirac point.
From Fig.~\ref{fig:split} it is clear that the two phonon modes with
$q_0$ and $q_\pi$ have different phonon energies, which results in
the double peak structure of the $D$ band.
The difference between the energy of the phonon mode with
$q_0$ and that with $q_\pi$ is approximated by
\begin{align}
 \Delta \omega_D = \frac{\partial \omega_D}{\partial q}
 \Delta q,
 \label{eq:dome_D}
\end{align}
where $\partial \omega_D/\partial q$
is the slope of the phonon energy dispersion.
We interpret the dispersive behavior that occurs 
as a result of the dispersion relation of the phonon mode.
Then we have
\begin{align}
 \frac{\partial \omega_D}{\partial q}=
 \frac{\partial E_L}{\partial q}
 \frac{\partial \omega_D}{\partial E_L}.
 \label{eq:disperivep}
\end{align}
Putting $|\partial E_L/\partial q| =\hbar v_{\rm F}$
in Eq.~(\ref{eq:disperivep}), and 
combining it with Eqs.~(\ref{eq:dtheta_p}) and Eq.~(\ref{eq:dome_D}),
we obtain Eq.~(\ref{eq:disp}).
Actual splitting can be smaller than Eq.~(\ref{eq:disp})
due to several factors, such as (i) 
the phonon dispersion relation for the $D$ band 
not being exactly isotropic around the Dirac point,~\cite{grueneis09}
(ii) $\Theta$ is not exactly limited by $0$ or $\pi$.

%%%%%%%%%%%%%%%%%%%%%%%%%%%%%
%%%%%%%%%%%%%%%%%%%%%%%%%%%%%

\section{Prospects}\label{sec:out}

The $D$ band has a close relationship with the energy gap.
A direct relationship between the zone-boundary $A_{1g}$ mode 
and the energy gap is seen in the Pierls instability at armchair
nanoribbons~\cite{fujita97,son06_energ}.
In the Dirac equation, an energy gap is represented by 
a Dirac mass term that appears in the off-diagonal terms as
\begin{align}
 \hat{H} =
 \begin{pmatrix}
  v_{\rm F}\bsigma \cdot \hat{\bf p} & m \sigma_x \cr
  m^* \sigma_x & v_{\rm F} \bsigma' \cdot \hat{\bf p}
 \end{pmatrix}.
\end{align}
The spectrum has an energy gap $2|m|$, 
because the energy dispersion is given by $\pm \sqrt{(\hbar v_{\rm F}
k)^2 +|m|^2}$.
The zone-boundary $A_{1g}$ mode is related to the Dirac mass because
the electron-phonon interaction [Eq.~(\ref{eq:HDeff})] 
appears as a mass term in the Dirac equation:
$m\to g_{\rm ep}e^{-i{\bf q}\cdot {\bf r}}$.
The dominance of intervalley backward scattering
becomes clear with respect to the mass term,
because a nonzero mass tends to stop a massless particle by backward
scattering.

Interestingly, the armchair edge itself can be modeled as a singular mass term
in the Dirac equation, whereby Eq.~(\ref{eq:armDirac}) is obtained as a
solution of the model~\cite{sasaki10-chiral}.
This theoretical framework leads us to find 
a relationship between the dominance of intervalley
backward scattering represented by Eq.~(\ref{eq:mateleD}) and BZF.
To see a connection between them, 
it is important to recognize that in deriving Eq.~(\ref{eq:mateleD}), 
the orbital structures of
Eqs.~(\ref{eq:wf1}) and (\ref{eq:wf2}) play a decisive role.
These orbitals have to relate to each other by mirror symmetry
with respect to the $x$-axis [$x\to -x$ or $k_x \to -k_x$] in Fig.~\ref{fig:unit}.
In fact, Eq.~(\ref{eq:wf2}) is constructed by replacing $\Theta$ with
$\pi-\Theta'$ in Eq.~(\ref{eq:wf1}).
The orbital should be invariant under this replacement.

Defects, such as a lattice vacancy and a topological defect, 
are considered to be sources that increase the $D$ band intensity
[$c$-term in Eq.~(\ref{eq:geneformu})].
This speculation is reasonable, 
because creating a lattice vacancy inevitably involves the antibonding
orbital.
Unfortunately, the wave functions and the corresponding BZ 
in the presence of defects are difficult to construct 
rigidly in the framework of a tight-binding lattice model.
This difficulty prevents us from calculating the electron-phonon matrix
element exactly.
However, there is the possibility of obtaining a good
approximation of the matrix element using the Dirac equation.~\cite{ferreira11,basko08_reson,bena08,novikov07,peres07}
%The $D$ band polarization behavior consists only of the electrons with
%bonding or antibonding orbitals.
%It is also meaningful to note that for zigzag edge, our analysis 
%shows that the BZ holding is not happening.
%This is because of the pseudospin flips at the zigzag edge, on the
%contrary to that at armchair edge.
%Our result suggests that the physical mechanism of the $D$ band is
%important in controlling carbon network by laser light~\cite{begliarbekov11}.

\section{Conclusion}\label{sec:con}

An orbital and wave vector are the basic idea for a molecular and a crystal, respectively.
Knowing the correlation between them is the key to understand the $D$ band.
The $D$ band originates from two orbitals: 
the bonding and antibonding orbitals on the $k_x$-axis.
Observing the $D$ band is the same thing as 
selecting the two orbitals from the various orbitals that compose the iso-energy section of the Dirac cone.
This idea leads us to expect the optical control of edge chiralities~\cite{begliarbekov11}.
A slight asymmetry between the bonding and antibonding orbitals,
induced by the trigonal warping effect, may be important in terms of
understanding the stability of the armchair edge under laser light irradiation.
A closer study of the $D$ band based on the Dirac equation will be
fruitful for a further investigation of the physics of the $D$ band.

\section*{Acknowledgments}

We were informed that in the case of two-phonons Raman
scattering the group velocities of the electron before and
after the phonon emission turn out to be anti-parallel with a much higher 
precision than prescribed by Eq.~(\ref{eq:mateleD}).
This has been discussed by Basko in Sec.VI.A of~\cite{basko08_theor}, 
and by Venezuela in Sec.III.E.2 of~\cite{venezuela11}.

\appendix

\section{Resonance condition}\label{app:rc}

Here, we show that 
the application of resonance condition is validated
unless the mean lifetime of intermediate state is extremely short.

The probability amplitude of a Raman process that contributes to the $D$
band is written as 
$\langle {\bf A}_{\rm sc}, D({\bf q}) | {\bf A}_{\rm in}\rangle$,
where ${\bf A}_{\rm in}$ (${\bf A}_{\rm sc}$) is the vector potential
for an incident (scattered) light and $D({\bf q})$ represents an excited $A_{1g}$
mode with wave vector ${\bf q}$.
We employ perturbation theory for obtaining the corresponding matrix
element:
\begin{align}
 M= (ev_{\rm F})^2 g_{\rm ep} \int_{-\infty}^{\infty} d\varepsilon {\rm Tr} \left[
 G_0(\varepsilon-E_L)
 \left( \bsigma\cdot {\bf A}_{\rm sc} \right)
 G_0(\varepsilon-\hbar \omega_{D}) \left( \sigma_x
 \right) G_0(\varepsilon) \left( \bsigma\cdot {\bf A}_{\rm in} \right) 
 \right],
 \label{eq:sma}
\end{align}
where $G_0(\varepsilon)$ is the propagator defined by
\begin{align}
 G_0(\varepsilon) = 
 \sum_{s=\pm 1} \sum_{\bf k} \frac{\psi^s({\bf k}) \psi^s({\bf k})^\dagger}{
 \varepsilon-sE_{k}+i\gamma},
 \label{eq:pro}
\end{align}
with $E_k = \hbar v_{\rm F}|{\bf k}|$,
$\hbar/\gamma$ is the mean lifetime of an intermediate state, and $\psi^s({\bf k})$ is defined
by the right-hand side of Eq.~(\ref{eq:wfK}).
By putting Eq.~(\ref{eq:pro}) into Eq.~(\ref{eq:sma}), 
we can find that $M$ leads to 
\begin{align}
 M \propto \int_{-\infty}^{\infty} d\varepsilon {\rm Tr} \left[
 \sum_{s,{\bf k}}
 \frac{\psi^{-s}({\bf k})\psi^{-s}({\bf k})^\dagger  \left( \bsigma\cdot
 {\bf A}_{\rm out} \right) \psi^s({\bf k}) \psi^s({\bf k})^\dagger
 \sigma_x \psi^s({\bf k}) \psi^s({\bf
 k})^\dagger \left( \bsigma\cdot {\bf A}_{\rm in}
 \right)}{(\varepsilon-E_L+sE_{k}+i\gamma)(\varepsilon-\hbar
 \omega_{D}-sE_{k}+i\gamma)(\varepsilon-sE_{k}+i\gamma)} 
 \right].
 \label{eq:Mat}
\end{align}
In deriving Eq.~(\ref{eq:Mat}) from Eq.~(\ref{eq:sma}),
we used the fact that the $D$ band is described as a
first-order process in the folded BZ, that is, 
the wave vector of a photo-excited electron (or hole) does not change
when it emits an $A_{1g}$ mode.
Furthermore, since the wavelength of a laser light is much longer than the
characteristic length scale of the electron,
there is no change of the wave vector ${\bf k}$ of an electron (or hole)
throughout the process.
The amplitude $M$ is obtained by employing the summation over ${\bf k}$ 
[$\sum_{\bf k}$ in Eq.~(\ref{eq:Mat})]
that satisfy the momentum conservation given by
Eq.~(\ref{eq:phwavevec}), $E_k \cos\Theta={\rm constant}$.

Since $\psi^s({\bf k})$ is independent of $E_k$, 
the numerator of Eq.~(\ref{eq:Mat}) does not change when the value of $E_k$ changes.
However, due to the constraint $E_k \cos\Theta={\rm constant}$,
we have to sum over $E_k$ 
while taking into account a change of $\Theta$ that determines the numerator. 
As a result, there is a possibility that 
the polarization dependence of the $D$ band is sensitive to a change of $E_k$.
Let $f(\varepsilon,E_{k})$ be the denominator of Eq.~(\ref{eq:Mat})
with $s=1$;
\begin{align}
 f(\varepsilon,E_{k})= \frac{1}{(\varepsilon- E_L+E_{k}+i\gamma)(\varepsilon-\hbar \omega_{D}-E_{k}+i\gamma)(\varepsilon-E_{k}+i\gamma)}.
\end{align}
There are two resonance conditions, $\varepsilon = E_L/2$ and
$(E_L+\hbar\omega_D)/2$.\footnote{The second resonance condition is for
anti-Stokes process which is negligible at room temperature.}
When $\varepsilon = E_L/2 \equiv \varepsilon_R$ 
(i.e., the first resonance condition is satisfied), the function becomes
\begin{align}
 f(\varepsilon_{R},E_k) = 
 \frac{1}{(\varepsilon_{R}-\hbar \omega_D-E_{k}+i\gamma)}
 \frac{1}{(E_{k}-\varepsilon_{R})^2+\gamma^2}.
 \label{eq:fn}
\end{align}
A strong resonance appears for $E_k = \varepsilon_R$ when $\gamma$ is sufficiently small.
In this case, for each final state defined by the phonon momentum 
it is sufficient to pick one intermediate state that
satisfies the resonant condition $E_k = \varepsilon_R$ 
and the momentum conservation $E_k \cos\Theta=\varepsilon_R
\cos\Theta_R$, and then just sum the square of the resulting Raman
matrix element over the final states, as we have done in
Sec.~\ref{ssec:fin}.

To see the effect of off-resonant intermediate states 
($E_k \ne \varepsilon_{R}$), 
we perform the integral over $E_k$ as
\begin{align}
 M \propto \int_0^\infty  g(\cos\Theta,\sin\Theta) \cos\Theta
 f(\varepsilon_R,E_k)E_k dE_k,
\end{align}
where $g(\cos\Theta,\sin\Theta)$ denotes the optical matrix element in
the numerator, and $\cos \Theta$ originates from the electron-phonon matrix element.
By changing the variable $E_k$ to $\theta$ via
$E_k-\varepsilon_{R}=\gamma \tan\theta$, we have
\begin{align}
 M \propto -\int_{-\frac{\pi}{2}}^{\frac{\pi}{2}}
 g\left( \frac{\cos\Theta_R}{1+\frac{\gamma}{\varepsilon_R}\tan\theta},
 \pm \sqrt{ 1-\left(
 \frac{\cos\Theta_R}{1+\frac{\gamma}{\varepsilon_R}\tan\theta} \right)^2
 }\right)
 \frac{\varepsilon_R \cos\Theta_R}{(\hbar \omega_D + \gamma \tan\theta -i\gamma)\gamma}d\theta. 
\end{align}
In Raman spectroscopy, $\varepsilon_{R}=1$ eV is a typical value
and $\gamma /\varepsilon_{R} \ll 1$ is satisfied in general.
Note that $(\gamma/\varepsilon_R) \tan\theta$ is enhanced in the $\theta
\to \pm \pi/2$ limits, which could modify the light polarization
dependence of the $D$ band.
However, the denominator is also enhanced in these limits, and the
contribution of these intermediate states to $M$ is suppressed overall. 
Thus, we neglect the $\theta$ dependence of the numerator $g$, and get
\begin{align}
 M \propto -g\left( \cos\Theta_R, \sin\Theta_R \right)\cos\Theta_R
 \int_{-\frac{\pi}{2}}^{\frac{\pi}{2}} 
 \frac{\varepsilon_R }{(\hbar \omega_D + \gamma \tan\theta -i\gamma)\gamma}d\theta. 
\end{align}
Since the integral is independent of $\Theta_R$,\footnote{
$ \int_{-\frac{\pi}{2}}^{\frac{\pi}{2}} 
 \frac{\varepsilon_R }{(\hbar \omega_D + \gamma \tan\theta
 -i\gamma)\gamma}d\theta=\pi
\frac{\varepsilon_R }{\gamma }\frac{1-i\frac{\gamma}{\hbar \omega_D}}{\hbar \omega_D-2i\gamma}$.
}
we conclude that the polarization behavior of the $D$ band 
(such as the depolarization ratio) is insensitive to 
the existence of off-resonant intermediate states 
unless $\gamma$ is comparable to $\varepsilon_R$.
The maximum for the $D$ band intensity depends on the $\gamma$ value, of course.
Our conclusion does not change
even if the integral over $\varepsilon$ is taken into account, 
because the integral is taken into account by replacing $\gamma$ with 
$\gamma -i\delta\varepsilon$ (where $\varepsilon =\varepsilon_R + \delta
\varepsilon$) in Eq.~(\ref{eq:fn}).

\bibliographystyle{mdpi}
\makeatletter
\renewcommand\@biblabel[1]{#1. }
\makeatother

% \bibliography{/Users/sasakikenichi/bib/sasaki,/Users/sasakikenichi/bib/sasaki_mgm}

\end{document}